# Bright Coherent Ultrahigh Harmonics in the keV X-Ray Regime from Mid-Infrared Femtosecond Lasers

Tenio Popmintchev[1*], Ming-Chang Chen[1], Dimitar Popmintchev[1], Paul Arpin[1], Susannah Brown[1], Skirmantas Ališauskas[2], Giedrius Andriukaitis[2], Tadas Balčiunas[2], Oliver Mücke[2], Audrius Pugzlys[2], Andrius Baltuška[2], Bonggu Shim[3], Samuel E. Schrauth[3], Alexander Gaeta[3], Carlos Hernández-García[4], Luis Plaja[4], Andreas Becker[1], Agnieszka Jaron-Becker[1], Margaret M. Murnane[1], and Henry C. Kapteyn[1]

[1]*JILA, University of Colorado at Boulder, Boulder, CO 80309, USA*
[2] *Photonics Institute, Vienna University of Technology, Vienna A-1040, Austria*
[3]*School of Applied and Engineering Physics, Cornell University, Ithaca, NY 14853, USA*
[4]*Grupo de Investigación en Óptica Extrema, Universidad de Salamanca, Salamanca E37008, Spain*
* Ph: 1.303.807.7276; E-mail: popmintchev@jila.colorado.edu

## Abstract

High harmonic generation traditionally combines ~100 near-infrared laser photons, to generate bright, phase matched, extreme ultraviolet beams when the emission from many atoms adds constructively. Here we show that by guiding a mid-infrared femtosecond laser in a high pressure gas, ultrahigh harmonics can be generated up to orders > 5000, that emerge as a bright supercontinuum that spans the entire electromagnetic spectrum from the ultraviolet to > 1.6 keV, allowing in-principle the generation of pulses as short as 2.5 attoseconds. The multi-atmosphere gas pressures required for bright, phase matched emission also supports laser beam self-confinement, further enhancing the x-ray yield. Finally, the x-ray beam exhibits high spatial coherence, even though at high gas density, the recolliding electrons responsible for high harmonic generation encounter other atoms during the emission process.

The unique ability of x-rays to capture structure and dynamics at the nanoscale has spurred the development of large-scale x-ray free-electron lasers based on accelerator physics, as well as high harmonic generation (HHG) techniques in the x-ray region that employ tabletop femtosecond lasers. The HHG process represents nonlinear optics at an extreme, enabling unprecedented femtosecond-to-attosecond duration pulses with full spatial coherence,(*1-6*) which make it possible to track the dynamics of electrons in atoms, molecules and materials.(*7-12*) X-rays can probe the oxidation or spin state in molecules and materials with element-specificity, because the position of the characteristic x-ray absorption edges of individual elements is sensitive to the local environment and structure. Ultrashort x-ray pulses can capture the coupled motions of charges, spins, atoms and phonons by monitoring changes in absorption or reflection that occur near these edges as a material or molecule changes state or shape. However, many inner-shell absorption edges in advanced correlated-electron, magnetic and catalytic materials (Fe, Co, Ni, Cu) lie at photon energies nearing 1 keV.(*13-15*) In contrast, most applications that use HHG light have been limited to the extreme ultraviolet (EUV) region of the spectrum (< 150 eV), where efficient frequency upconversion is possible using widely available Ti:Sapphire lasers operating at 0.8 μm wavelength. We therefore sought to extend bright HHG to a higher energy soft x-ray region.

High harmonic generation is a universal response of atoms and molecules in strong femtosecond laser fields.(*16, 17*) In a simple analogy, HHG represents the coherent version of the Röntgen X-ray tube: instead of boiling electrons off a hot filament, accelerating them in an electric field, and generating incoherent x-rays when the high-energy electrons strike a target, HHG begins with tunnel ionization of an atom in a strong laser field. The portion of the electron wavefunction that escapes the atom is accelerated by the laser electric field, and when driven back to its parent ion by the laser, can coherently convert its kinetic energy into a high harmonic photon. The highest-energy HHG photon emitted is given by the microscopic single-atom cutoff rule: $h\nu_{SA\ cutoff}=I_p+3.17U_p$, where $I_p$ is the ionization potential of the gas, and $U_p \propto I_L\lambda_L^2$ is the quiver energy of the liberated electron in a laser field of intensity $I_L$ and wavelength $\lambda_L$.

Generating bright, fully coherent HHG beams requires macroscopic phase matching,(*18*) wherein the laser and high-order nonlinear polarization propagate in phase (at the speed of light *c*) throughout a medium to ensure that the HHG light emitted from many atoms adds coherently.(*1, 19, 20*) Phase matching is achieved by balancing the neutral gas and free-electron plasma dispersion experienced by the laser, and is only possible up to some critical ionization level that depends on the gas species and laser wavelength (Fig. S1). Any geometric contributions to the laser propagation must also be considered (see supporting text). Because ionization increases with laser intensity, the critical ionization limits the highest photon energy for which phase matching can be implemented. Recent work explored the wavelength dependence of the HHG yield,(*21-24*) which scales as $h\nu_{PM\ cutoff} \propto \lambda_L^{1.7}$ under phase matched conditions.(*25-27*) Using 2 µm lasers (0.62 eV photons) to drive HHG, bright harmonics extend to > 0.5 keV, (*26*) demonstrating phase matching of an >800 order nonlinear process (note that only odd order harmonics are emitted to conserve angular momentum).

In this work, bright high harmonic x-ray supercontinua with photon energies spanning from the EUV to 1.6 keV (<7.7 Å) are generated by focusing 3.9 µm wavelength pulses from a tabletop femtosecond laser into a waveguide filled with He gas (see Fig. 1). The multi-atmosphere pressures necessary for efficient x-ray generation also support laser beam self-confinement, enhancing the x-ray yield by another order of magnitude. We observe coherent, laser-like, x-ray beams, despite the fact that ultrahigh harmonic generation occurs in a regime where the laser-driven electrons encounter many neighboring atoms before they re-encounter their parent ions. Our calculations indicate that the keV-bandwidth coherent supercontinuum has a well behaved chirp that, when compensated, could support a single-x-ray-cycle 2.5 attosecond pulse duration. Finally, we show that in the keV region, a much higher order nonlinear process is required for phase matching than is required for harmonic emission from a single atom.

In our experiment, 6-cycle FWHM (80 fs), 10 mJ pulses, centered at 3.9 µm wavelength are generated at 20 Hz as the idler output of an optical parametric chirped-pulse amplification (OPCPA) laser system.(*28, 29*) X-rays are generated by focusing the laser beam into a 200 µm diameter, 5 cm long, gas-filled hollow waveguide capable of sustaining pressures of up to 80 atm in a differentially-pumped geometry. The HHG spectrum is then captured using a soft x-ray spectrometer and x-ray CCD camera. Figure 1B shows the phase matched HHG emission from He, which extends to >1.6 keV (<7.7 Å). The phase-matched HHG cutoff energy agrees well with numerical predictions plotted in Fig. 2A for 3.9 µm driving lasers i.e. $h\nu_{PM\ cutoff} \propto \lambda_L^{1.7}$.(*25-27*) This bright x-ray supercontinuum is ideal for x-ray spectroscopy measurements, spanning multiple inner-shell absorption edges simultaneously (Figs. 1B and S2), as has already been demonstrated in the EUV region for HHG driven by multi-cycle 0.8 µm lasers where a quasi-continuous HHG spectrum is emitted.(*15, 30, 31*)

The x-ray flux from He scales quadratically with pressure (number of emitters), as shown in Fig. 1A, reaching a maximum at very high gas pressures of ~ 35 atm, where both phase matching and laser beam self-confinement are optimized. At higher pressures, the x-ray flux decreases due to reabsorption of the generated harmonics by the high-pressure gas, and because of energy loss experienced by the laser when coupling into the waveguide. Microscopically, quantum diffusion leads to spreading of the electron wavepacket, decreasing the recombination probability and thus the single-atom HHG yield,(*22, 23, 24*) which scales with the laser wavelength as $\sim \lambda_L^{-6.5}$. Specifically, the single atom HHG yield is ~$3\times10^5$ smaller at 3.9 μm compared with 0.8 μm. Fortunately, the low single-atom yield can be compensated by coherently combining HHG from a large number of emitters (high gas density and medium length), which is possible in part because the gas becomes increasingly transparent at photon energies approaching the keV region. The HHG signal builds up over a density-length product comparable to the absorption depth of the x-ray light, leading to nearly constant brightness of the HHG emission from 0.3 to 1 keV. An approximate brightness of $10^5$ photons/shot (corresponding to $10^6$ photons/s at 20 Hz) is observed in a fractional bandwidth of 1% at 1 keV. Past work successfully used 0.8 μm lasers to demonstrate keV harmonics using a ~1000 order nonlinear process, but with much reduced flux (4 to 5 orders of magnitude lower) because phase matching is not possible in this regime.(*14*) Thus, surprisingly, for macroscopic phase matching, the required harmonic order is much higher than that required to generate the same photon energy from a single atom using shorter laser wavelength.

We can now present a unified picture for phase matched high harmonic upconversion, spanning the electromagnetic spectrum from the VUV to > keV x-ray photon energies, that includes both the microscopic and macroscopic physics. To validate theory, we tuned our driving laser to different wavelengths from the UV to the mid-IR, and then implemented pressure-tuned phase matching to optimize the HHG flux at each laser wavelength with the optimal laser intensities dictated by the critical ionization of the medium (Fig. 2 and supporting text). The required optimal pressures and interaction lengths evolve from < 0.1 atm and a few mm in the VUV region, to tens of atm and multi-cm lengths in the x-ray region. Figures 2A and C plot the optimized phase matched cutoffs and spectra for different driving laser wavelengths. To efficiently generate high harmonics, the order of the nonlinearity must increase from ~11 in the vacuum UV, to >5000 in the keV region. This represents an extreme both in terms of the order of a nonlinear process, and for phase matching. The bright phase matched HHG spectra evolve from a single harmonic in the VUV, into a broad x-ray supercontinuum spanning thousands of harmonics in the soft x-ray region. Phase matching shuts off in the VUV, at energies near the ionization potential of the nonlinear gas medium, as the HHG and driving laser wavelengths converge (see Fig. 2A). The phase-matched HHG conversion efficiencies reach $10^{-3}$ to $10^{-4}$ in the VUV region, compared with $10^{-5}$ in the EUV using 0.8 µm lasers, and $10^{-6}$ to $10^{-7}$ in the x-ray region. Moreover, in the VUV region, phase matching occurs at relatively high levels of ionization of tens of percent (Fig. S1).

Remarkably, tunnel ionization of the atomic gas medium dominates in all phase matching regimes. When driven by UV light, the effective potential (which is a superposition of the Coulomb and laser fields) oscillates rapidly, allowing a very short time interval for the electron to tunnel. However, the required laser intensity for HHG is extremely high (e.g. >$10^{15}$ W/cm$^2$), so that tunnel ionization is more probable than multiphoton ionization. For mid-IR laser wavelengths, the slowly oscillating effective Coulomb potential can be considered quasi-static. Therefore, although the laser intensity decreases to maintain phase matching, tunnel ionization is still more probable than multiphoton ionization. Because the physics of ionization does not change, we can use an analytical description of tunneling (Amossov-Delone-Krainov model (*32*)) to derive a generalized analytic HHG phase matching cutoff rule (Eq. 1), validated by comparing with experiment, as well as numerical and quantum theory (see supporting text):

$$h\nu_{PM\ cutoff} = I_p + \frac{\alpha I_p^3}{\ln^2\left[\frac{\beta I_p \tau_L}{-\ln(1-\eta_{CR}(\lambda_L))}\right]} \lambda_L^2 \qquad (1)$$

Here α and β are constants that depend on the laser pulse shape and the state from which the electron is tunnel ionized. This analytical expression gives some physical insight into phase matching of the HHG upconversion process. The small deviation of $\lambda_L^{(1.5-1.7)}$ from the $\lambda_L^2$ scaling of the ponderomotive energy incorporates the proper scaling of the laser intensity and arises from the scaling of the critical ionization $\eta_{CR}$, which decreases by 4 orders of magnitude from the UV to mid-IR driving laser wavelengths. Short, few-cycle laser pulses make it possible to generate higher-energy photons before the critical ionization level is exceeded. However, this approach yields diminishing returns for pulses shorter than 5 to 10 cycles, and only leads to modest enhancements in HHG flux and phase matching cutoff. The most substantial HHG enhancement (by orders of magnitude) arises when the right combination of laser wavelength, gas pressure-length product and laser intensity is used.

Likewise, in contrast to conventional wisdom (see supporting text), Helium is generally the best atomic medium for harmonic generation, because of the absence of inner-shell absorption.(*25*) The absorption limit for HHG emission can be clearly seen in Fig. 2B, which plots the phase matched HHG emission from Ar, $N_2$ and Ne when driven by 3.9 μm light. There is a sharp drop-off in signal at the inner-shell absorption edges at 0.25 keV (Ar), 0.41 keV ($N_2$) and 0.87 keV (Ne), and therefore the true phase matching cutoff cannot be observed: without absorption, the phase matching limits would be ~0.5 keV (Ar, $N_2$) and ~1 keV (Ne).

Generating bright keV harmonics from atoms driven by mid-IR femtosecond lasers takes advantage of a remarkable convergence of favorable physics. First, the very high gas density required puts these experiments in a distinct regime of HHG from *non-isolated* emitters: spread of the ionized electron quantum wavepacket over its few-femtosecond free trajectory means that the electron will encounter many neighboring atoms. This contrasts with emission from dilute, isolated atoms for UV or EUV harmonic generation. As shown in Fig. S1, for keV harmonics the electron wave function in the continuum extends to ~ 500 Å, whereas the separation between the He atoms is ~15 Å at 10 atm pressure. However, the ionization levels are low at ~0.03%. For VUV/EUV harmonics, the electron typically extends ~ 2 to 20 Å between ionization and recollision, while the separation between atoms is ~ 70 Å at ~0.1 atm pressure, and phase matching occurs at ~10% ionization levels. Thus, HHG driven by mid-IR pulses liberates a thousandth as much of the electron wavefunction into the continuum compared to visible driving lasers, though it is spread over a 100x larger distance. Fortunately, our experimental results indicate that rescattering of this large and diffuse recolliding electron wavepacket from other atoms seems not to adversely influence the coherence of the emission, likely because the medium is weakly ionized. Evidence for this includes the well-formed, spatially coherent, x-ray beams (Fig. 4) and the remarkable quadratic growth (Fig. 1A) that continues from 0.2 atm (when the rescattering electron wavepacket can begin to encounter neighboring atoms) to more than 2 orders of magnitude higher pressure.

In a second extremely favorable convergence of extreme nonlinear optics, the multi-atmosphere gas pressures required for phase matched x-ray generation also overlaps with the parameter range where laser beam self-confinement is possible. Figure 3A plots the experimental x-ray emission from Ar driven by 3.9 μm lasers. The predicted phase-matching pressure is ~3 atm, and indeed we observe a peak in x-ray emission at that pressure. However, as the pressure is further increased, the x-ray yield first decreases and then increases quadratically, exhibiting a large enhancement at a pressure of 26 atm (about a factor of 10 when integrated over all soft x-ray HHG). The measured x-ray beam profile also dramatically narrows as the gas pressure increases (Fig. 3B), indicative of self-confinement of the driving laser. Essentially, the x-ray

HHG beam, imaged at the exit of the fiber, shrinks to less than a third of its former diameter while the x-ray signal increases tenfold (integrated over all orders) at pressures 7 times greater than required for phase matching.

To explore theoretically how macroscopic nonlinear effects augment HHG phase matching, we numerically simulated nonlinear pulse propagation in a hollow waveguide filled with high-pressure gas by extending and expanding previous simulations to longer wavelengths.(*33, 34*) Our simulations show that as the gas pressure increases beyond that required for phase matching, the peak laser intensity is stabilized (see Figs. S3 and S4). We also observe strong spatio-temporal compression and localization of the driving laser during self-confinement due to Kerr effect and plasma generation, which also enhances the HHG yield. Figure 3B plots the calculated beam profiles at the phase matching (3.5 atm) and higher pressures (26 atm). A stable self-confined beam forms at the higher gas pressures and persists for centimeter distances. As discussed in the supporting text, we can observe experimentally and theoretically that self-confinement also enhances phase matching in other gases such as He (Fig. S4) and molecular $N_2$.

When phase matched, the spatial quality of the x-ray beam is excellent. Figure 4 shows the x-ray beam and the Young's double-slit diffraction patterns taken by illuminating 5 μm slits (separated by 10 μm) with an x-ray supercontinuum generated in He and Ne, spanning 7.7 to 43 Å and 14 to 43 Å, respectively. There is excellent agreement between the experimentally observed and theoretically predicted diffraction pattern. A plot of the expected diffraction pattern from incoherent x-ray illumination is also shown in Figs. 4B and C for the same experimental geometry, proving that the high fringe visibility is due not to the small pinhole size but rather to the high spatial coherence of the x-ray beam itself. This measurement is extremely challenging at short wavelengths: very small slit widths are required so that the light from each slit diffracts sufficiently to ensure overlap and interference at the detector (3.5 m away from the slits). Thus, the throughput is very small. This spatial coherence measurement clearly demonstrates that coherent diffractive imaging will be possible with near wavelength spatial resolution, as has been achieved using HHG beams and synchrotron sources in the EUV and soft x-ray regions.(*35, 36*)

To predict the temporal properties of the HHG radiation, we theoretically analyzed HHG driven by 1 and 6 cycle FWHM mid-IR laser pulses, with peak intensities of 4.1 and 3.3 x $10^{14}$ W/cm$^2$ respectively, from single atoms and also in a phase matched regime. Our calculations, based on the strong field approximation and discrete dipole approach,(*37*) confirm the femtoseocnd time scale of the x-ray bursts from a single atom and also after propagation (see supporting text). Our calculated phase matched HHG spectra agree well with those measured experimentally (Figs. 1B and C) and show that the HHG chirp is well behaved (Figs. 1C and S5) over the near-keV bandwidth that, when compressed, is sufficient to support a single-cycle, 2.5 attosecond pulse in the Fourier limit. For 3.9 μm driving lasers in the single-atom case, contributions from the short and long trajectories lead to a parabolic chirp, whereas after propagation, the phase matched short trajectory contribution leads to a positive, quasi-linear, chirp. The current limit of theory allows us to simulate HHG propagation over 20 μm distances at high pressures, and predicts that the uncompressed HHG temporal emission consists of a series of ~ 3 intense bursts of 1 to 3 fs duration, due to the very long 13 fs period of the multicycle 3.9 μm driving laser field (Figs. 1C and S5). However, for longer propagation distances, bright HHG emission in the form of a single isolated x-ray burst is likely expected. This is because phase matching is transient, and favors x-ray emission from a single half-cycle where the phase matching is optimal. This has been verified experimentally in the EUV - even without stabilizing the carrier wave with respect to the pulse envelope.(*25, 38*)

Experimental verification of these predictions will require the development of characterization methods that can sample ultrabroad bandwidth x-ray waveforms at different photon energies. This challenge is illustrated in Fig. 1B, where the narrow dip at 0.54 keV corresponds to oxygen K-edge absorption. It is not clear that any atomic or molecular system can interact with a keV bandwidth, because processes such as photoionization involve significantly slower timescales. However, the chirped x-ray supercontinua already represent a promising multiple-atomic-site probe with sub-femtosecond time resolution, analogous to the chirped white light (visible) continua used to probe many absorption features simultaneously, perfectly synchronized to the driving laser. Given our current experimental and theoretical findings, it may be possible to extend HHG to hard x-ray wavelengths and broader bandwidths, so that zeptosecond time scales may be accessible.

**FIGURE CAPTIONS**

**Figure 1. A.** Schematic illustration of the coherent keV x-ray supercontinua emitted when a mid-IR laser pulse is focused into a high-pressure gas-filled waveguide. The experimental phase-matched harmonic signal grows quadratically with pressure, demonstrating excellent phase-matched coherent buildup. **B.** Experimental HHG spectra emitted under full phase matching conditions as a function of driving laser wavelength (yellow: 0.8 µm, green: 1.3 µm, blue: 2 µm and purple: 3.9µm). Inset: Fourier transform-limited pulse duration of 2.5 attoseconds. **C.** Calculated spectrum and temporal structure of one of the phase matched HHG bursts driven by a 6 cycle FWHM, 3.9 µm pulse at a laser intensity of $I_L = 3.3 \times 10^{14}$ W/cm$^2$.

**Figure 2. A.** Predicted and observed HHG phase matching cutoffs as a function of laser wavelength from the UV to mid-IR. Solid circles show the observed cutoffs, open circles show the predicted cutoffs for Ar and Ne (which cannot be reached due to inner shell absorption, as shown in **B**). Solid squares on the left show the ionization potentials ($I_p$) of the different atoms. **C.** Unified picture of optimal phase-matched high harmonic upconversion, including microscopic and macroscopic effects.

**Figure 3. A.** Measured HHG yield in Ar as a function of pressure and photon energy, showing two peaks, one at 3 atm pressure due to pressure-tuned phase matching, and a second at 26 atm pressure due to the additional presence of laser beam spatio-temporal self-confinement. **B.** Experimental HHG beam profiles and calculated laser beam profiles after a propagation distance of 3.8 cm in the waveguide.

**Figure 4. A.** X-ray experimental beam profile. **B, C**. Young's double slit diffraction patterns taken by illuminating 5 µm slits, separated by 10 µm, with the beam shown in **A**. There is excellent agreement between the experimentally observed (purple line) and theoretically predicted (blue line) diffraction patterns. The broad bandwidth and very low divergence of the HHG beams limits the number of fringes observed. The expected diffraction assuming incoherent illumination is also given for comparison (black line), illustrating the high spatial coherence of the keV HHG source.

**Aknowledgements**

The experimental work was funded by a NSSEFF Fellowship, and the NSF Center for EUV Science and Technology. AG, AJB, MM, HK and Andreas Becker acknowledge support for theory from the US Air Force Office of Scientific Research (Grant no. FA9550-10-1-0561). Andreas Becker acknowledges support from U FP7 STREP #244068 and EuroStars Project #4335. Andrius Baltuska acknowledges support from Austrian Science Fund (FWF, Grant no. U33-16), and the Austrian Research Promotion Agency (FFG, Project 820831 UPLIT). CHG and LP acknowledge support from Junta de Castilla y León, Spanish MINECO (CSD2007-00013 and FIS2009-09522), and from Centro de Láseres Pulsados, CLPU. TP, MCC, Alon Bahabad, MM and HK have filed a patent on "Method for phase-matched generation of coherent soft and hard X-rays using IR lasers", US 61171783 (2008).

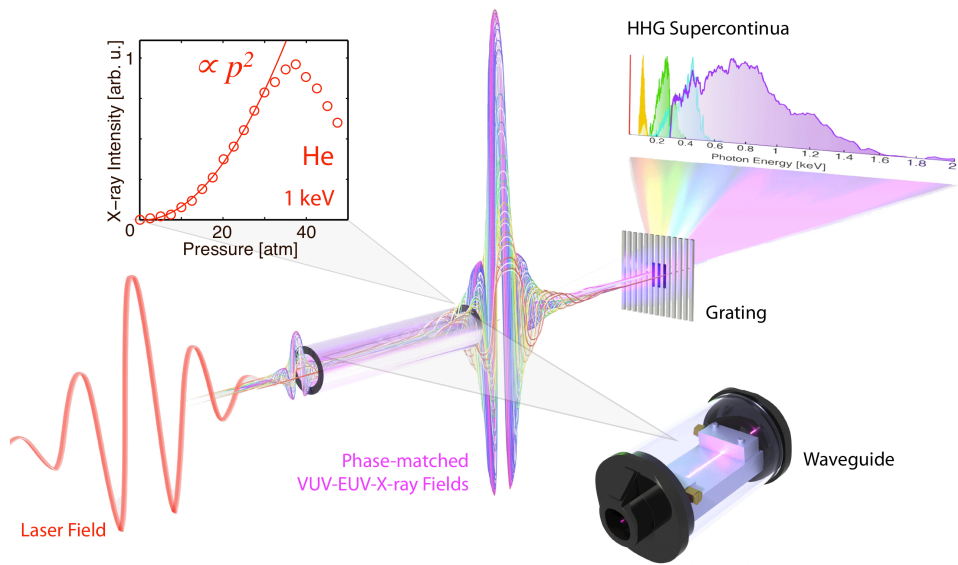

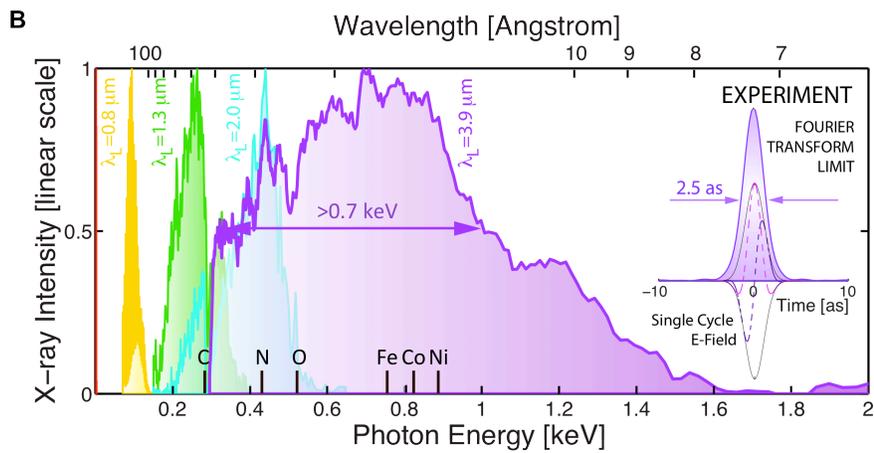

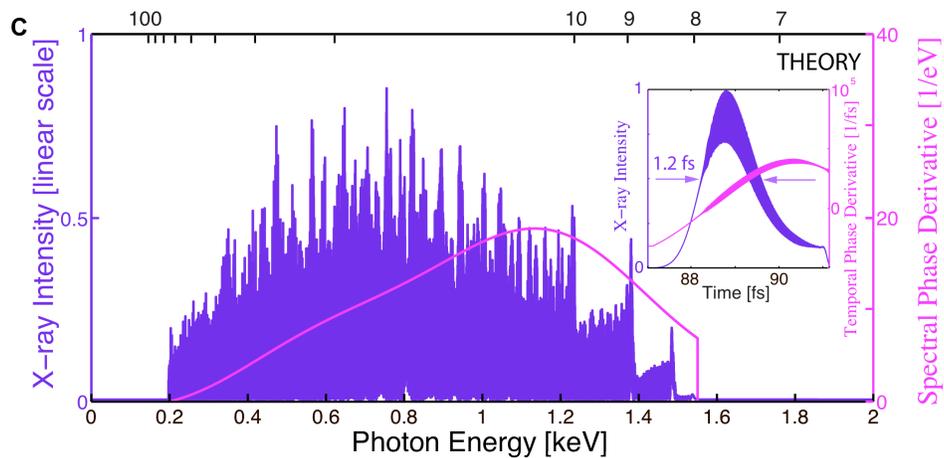

Figure 1.

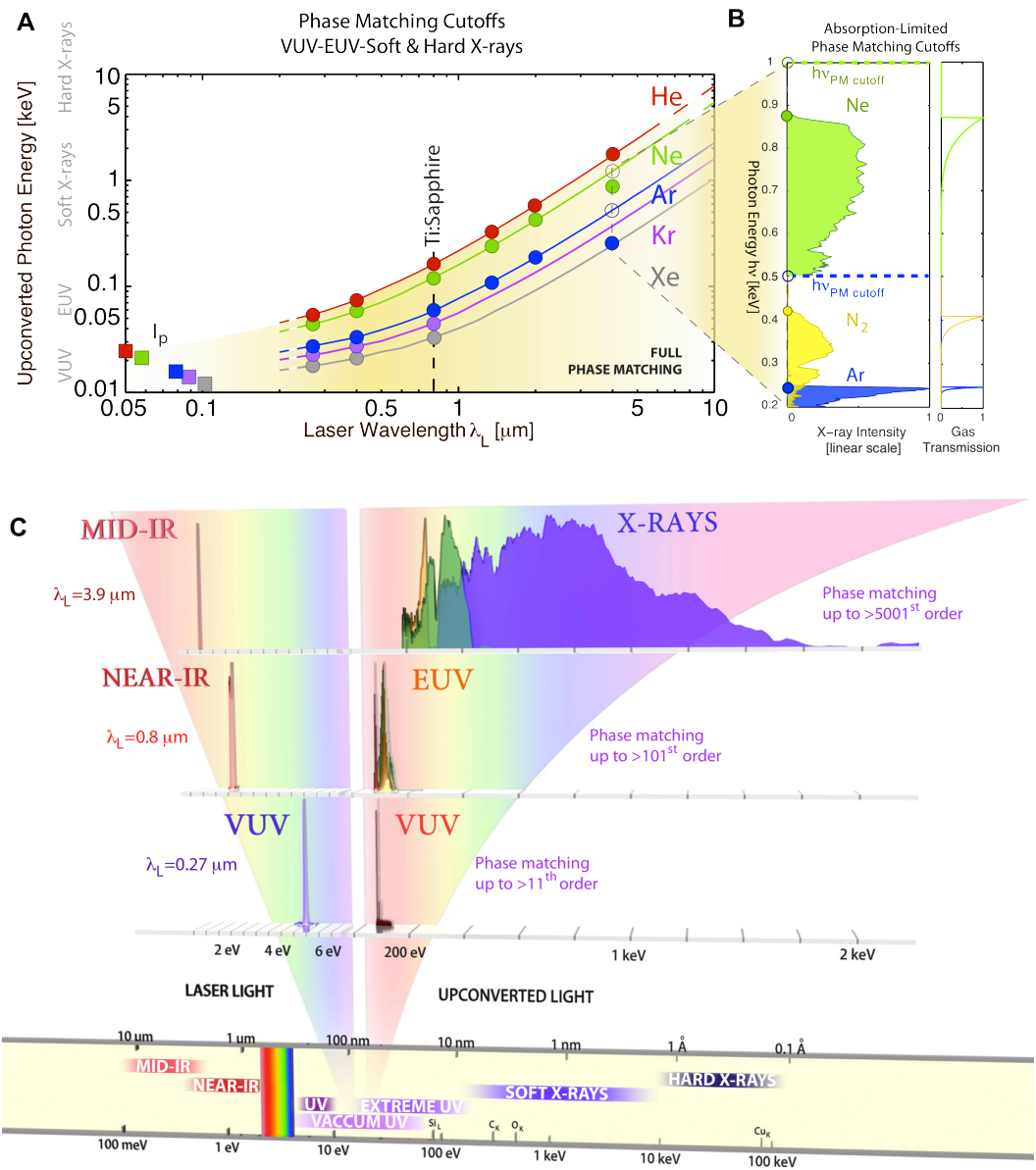

Figure 2.

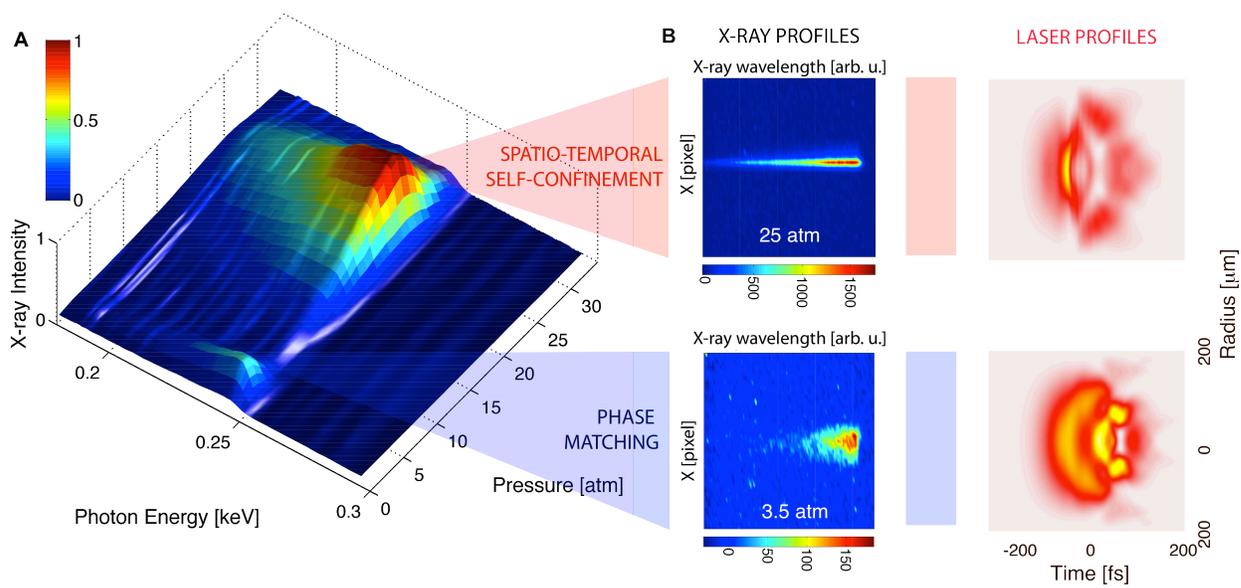

Figure 3.

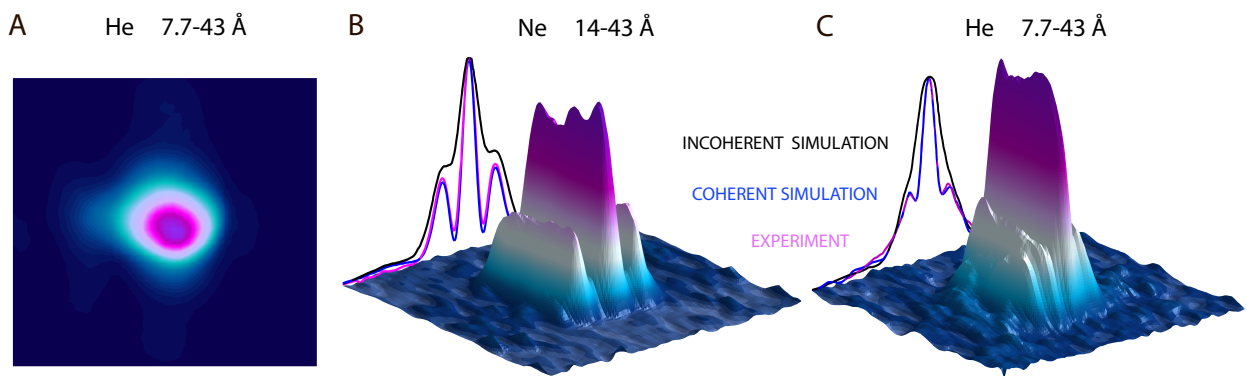

Figure 4.

# Bright Coherent Ultrahigh Harmonics in the keV X-Ray Regime from Mid-Infrared Femtosecond Lasers

Tenio Popmintchev[1*], Ming-Chang Chen[1], Dimitar Popmintchev[1], Paul Arpin[1], Susannah Brown[1], SkirmantasAlišauskas[2], Giedrius Andriukaitis[2], Tadas Balčiunas[2], Oliver Mücke[2], Audrius Pugzlys[2], Andrius Baltuška[2], Bonggu Shim[3], Samuel E. Schrauth[3], Alexander Gaeta[3], Carlos Hernández-García[4], Luis Plaja[4], Andreas Becker[1], Agnieszka Jaron-Becker[1], Margaret M. Murnane[1], and Henry C. Kapteyn[1]

[1]*JILA, University of Colorado at Boulder, Boulder, CO 80309, USA*
[2] *Photonics Institute, Vienna University of Technology, Vienna A-1040, Austria*
[3]*School of Applied and Engineering Physics, Cornell University, Ithaca, NY 14853, USA*
[4]*Grupo de Investigación en Óptica Extrema, Universidad de Salamanca, Salamanca E37008, Spain*
* Ph: 1.303.807.7276; E-mail: popmintchev@jila.colorado.edu

In this supporting text, we first provide additional details about the driving laser used to generate bright keV harmonics. Next, we expand on the discussion in the main text to highlight the differences between high harmonic generation in the EUV and keV regions of the spectrum. In the final sections, we explain our calculations of 1) mid-infrared laser propagation and self-confinement; 2) the high harmonic pulse duration and phase; and 3) the analytic formula for the scaling of phase matching in the VUV and x-ray regions of the spectrum.

**Femtosecond laser drivers at 3.9 μm, 0.4 μm and 0.27 μm wavelengths**

The mid-IR OPCPA laser system is based on a unique femtosecond Yb:$CaF_2$ chirped pulse amplifier which drives a cascaded optical parametric amplifier.(*22*) The subsequent stages of the OPCPA (based on KTA crystals) are pumped by a 20 Hz ps Nd:YAG laser system and produce record 35 mJ (uncompressed) and 10 mJ (compressed) pulse energies in the signal and the idler beams at 1.46 μm and 3.9 μm respectively. The 0.27 μm and 0.4 μm laser pulses were generated from the 3rd and 2nd harmonics of a high energy, 30 mJ, 25 fs, Ti:Sapphire laser system operating at a repetition rate of 10 Hz. To optimize phase-matched conversion into the VUV, pulse energies and durations of 4.5 mJ and 45 fs (3ω), and 8 mJ and 35 fs (2ω), were used.

**Bright keV ultrahigh harmonic generation**

The very high (multi atm) gas pressures required for optimum phase matching of high harmonic generation (HHG) at keV photon energies makes a differentially-pumped waveguide geometry the ideal target geometry.(*1*) Good differential pumping is required to avoid re-absorption of the generated x-rays by the gas, defocusing of the laser prior to entering the waveguide, and destructive interference in non-optimal pressure regions. The waveguide provides for minimum (through still substantial) gas load, since the gas flow and laser propagation apertures fully overlap. The waveguide designs incorporate a 1-4 cm constant-pressure section with 0.5 cm end sections for differential pumping; the shorter fiber lengths are used for the VUV HHG where the reabsorption of the upconverted light is stronger. As discussed in the main text, the very high gas densities required for phase matching of the HHG process in the keV region necessitates a new regime of HHG from *non-isolated* emitters in a gas; extreme quantum diffusion over several fs means that the electron wavefunction will encounter many neighboring atoms, ions, or electron wavepackets in the continuum, as illustrated in Fig. S1.

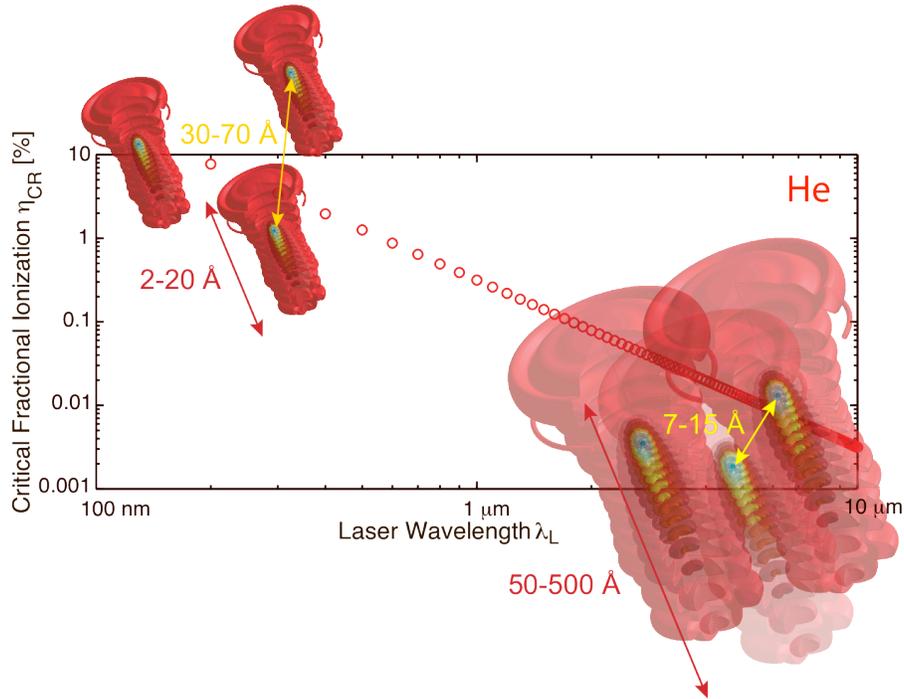

**Fig. S1.** The critical ionization fraction of He, above which phase matching is not possible, as a function of laser wavelength. Also shown are representations of the continuum electron wavepackets in the UV and keV regions.

Although not widely recognized previously, low-Z atoms (i.e. He) are nearly always optimum for HHG.(*29*) Past work dismissed HHG using He as exhibiting very low efficiency;

however, this simply resulted from the fact that wavelength and the intensity of the driving laser and the gas pressure were not well optimized. However, due to its relatively low index of refraction and absorption (especially in the soft x-ray region), the optimum pressure to maximize the harmonic yield using He is typically much higher than for other gasses, and virtually impossible to achieve in a gas-jet or even a cell geometry.

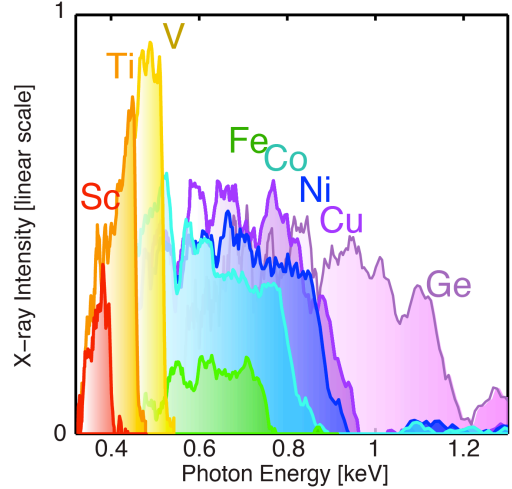

**Fig. S2.** The x-ray supercontinuum source enables site-specific spectroscopy at many L absorption edges simultaneously. Here thin metal films were placed between the HHG source and spectrometer.

Interestingly, even in the EUV region, when HHG is driven by 10-cycle FWHM 0.8 μm lasers, (where the laser photon energy is ~ 1.5 eV with individual harmonics 3 eV apart), there is measurable spectral amplitude between the well-separated HHG peaks, resulting in a quasi-continuum. This enables transient reflection or absorption experiments spanning multiple M-shell absorption edges. (*15, 24, 25*) For HHG driven by 3.9 μm lasers, the laser photon energy is 0.3 eV, while the separation between individual harmonics is 0.6 eV. Thus, due to the finite width of the harmonics, the x-ray HHG spectrum is continuous, spanning multiple L-shell absorption edges (see Fig. S2). The spectral resolution of the spectrometer, ~ 200 in the keV region, cannot distinguish the individual harmonic orders.

**Laser beam self-confinement in the mid-infrared**
To understand how nonlinear laser beam propagation and spatio-temporal confinement contributes to phase matching, we performed 3D simulations taking into account the spatial profile of the gas in the waveguide. The gas has a uniform pressure section in the center of the waveguide, while the entrance and exit sections (~ 0.5 cm long) have a density gradient. We therefore start the field propagation from a position outside the waveguide where the density of the gas is low. For propagation in the region before the waveguide, we use the nonlinear envelope equation (NEE) *(31)* given by:

$$\frac{\partial E}{\partial z} = \frac{i}{2k}\left(1+\frac{\partial}{\omega_0 \partial t}\right)^{-1}\nabla_\perp^2 E - i\frac{k''}{2}\frac{\partial^2 E}{\partial t^2} + \frac{ik}{n_0}\left(1+\frac{\partial}{\omega_0 \partial t}\right)\left(n_2|E|^2\right)E$$
$$-\frac{(\rho_n-\rho_e)W_{ADK}U}{2|E|^2}E - \frac{k}{2n_0^2\rho_c}\left(1+i\omega_0\tau_c\right)\frac{\omega_0\tau_c}{\left(1+\omega_0^2\tau_c^2\right)}\rho_e\left(E-\frac{(\partial E/\partial t)\tau_c}{1-i\omega_0\tau_c}\right)$$

where $E$ is the driving laser field, $z$ is the propagation distance, $k$ is the wavenumber corresponding to the central wavelength at 3.9 μm, $\omega_0$ is the angular frequency at the central wavelength, $k''$ is the group velocity dipersion parameter, $n_0$ is the refractive index of the gas, $n_2$ is the nonlinear index, $\rho_n$ is the neutral gas density, $\rho_e$ is the electron density, $W_{ADK}$ is the Ammosov-Delone-Krainov (ADK) ionization rate, $U$ is the first ionization energy, $\rho_c$ is the critical density at 3.9-μm wavelength, and $\tau_c$ is the electron collision time. The terms on the right side of the equation represents diffraction, dispersion, Kerr-nonlinearity, tunneling ionization absorption, and plasma defocusing and collisional absorption, respectively. We also assume that parameters are pressure dependent *(30)* and an exponential density gradient such that $\rho(z)=\rho_0\exp[-(z/L)^2]$, where $\rho_0$ is the neutral gas density at the entrance of the capillary. ADK tunneling ionization and collisional ionization are used for plasma generation. By numerically integrating the NEE, we determine the laser field profile at the input of the capillary.

For propagation inside the waveguide, we decompose the calculated input laser field into Bessel functions inside the capillary. Each Bessel mode amplitude is propagated independently, including modal dispersion and loss. After reconstructing the electric field from a superposition of the Bessel modes, we solve the nonlinear propagation in the spectral domain, including self-focusing, plasma defocusing, and absorption using the NEE. Since gases are injected through two holes that are located 5 mm from entrances of the capillary, there are also density gradients from the injection holes to two ends of the capillary. We use –

$$\rho(z)=\sqrt{\rho_0^2+(z/L)(\rho_{max}^2-\rho_0^2)}$$

where $\rho_{max}$ is the maximum gas density, which is proportional to the backing pressure.*(36)*

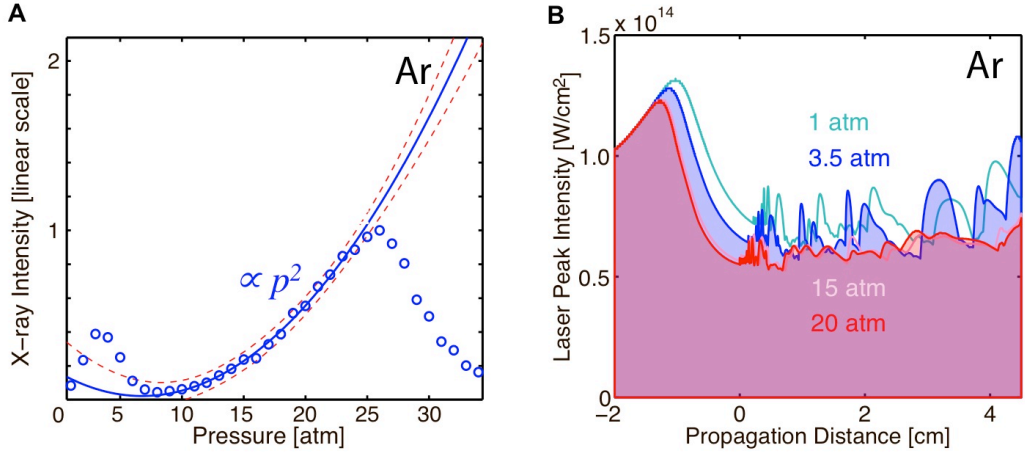

**Fig. S3. A**. Experimental x-ray yield as a function of pressure in Ar gas, showing two pressure peaks due to phase matching (3 atm) and laser beam self confinement (26 atm). **B**. Calculated on-axis peak laser intensity vs. propagation distance for different gas pressures. The capillary input is located at $z = 0$. A stable intensity is established due to spatio-temporal confinement at high > 20 atm pressures inside the waveguide.

Figures S3 B and S4 A plot the on-axis peak laser intensity as a function of propagation distance $z$, for Ar and He gas at different pressures (additional calculations for Ar are shown in Fig. 3 of the main text). For low pressures (~ 10 atm), the peak intensity inside capillary ($z > 0$) shows periodic oscillations due to energy exchange between the fundamental mode (i.e., lowest-order Bessel beam) and the higher-order modes. However, beam confinement due to the Kerr effect is observed at high pressures (> 15-20 atm for Ar and >30-40 atm for He) in the constant pressure region (0.5 cm $< z <$ 2.5 cm), since the critical power for self-focusing and filamentation is inversely proportional to gas pressure.(*37*) Here the focused beam $1/e^2$ radius at vacuum is 130 μm and the radius of the capillary is 200 μm. These findings match experiment very well.

The peak plasma density (Fig. S4 B) shows the same stabilization for higher pressures, showing that phase matching is supported by this laser beam self-confinement. The peak laser intensity and plasma density are in very good agreement with the experimental parameters for full phase matching, with minor deviations within the error of various parameters such as $n_2$, ionization models, and focus position with respect to the waveguide. Figure S4 C shows examples of spatio-temporal mode profiles for two different locations and pressures for He. Compared with mode fluctuations at lower pressures due to interference between the fundamental and the higher-order modes, the spatial modes at high pressures (at values of 40 atm close to the phase-matching pressure for λ=3.9 μm) are strongly localized near the waveguide center ($r = 0$). Strong spatio-temporal confinement is evident after the laser propagates 1.75 cm at high 40 atm pressure inside the waveguide.

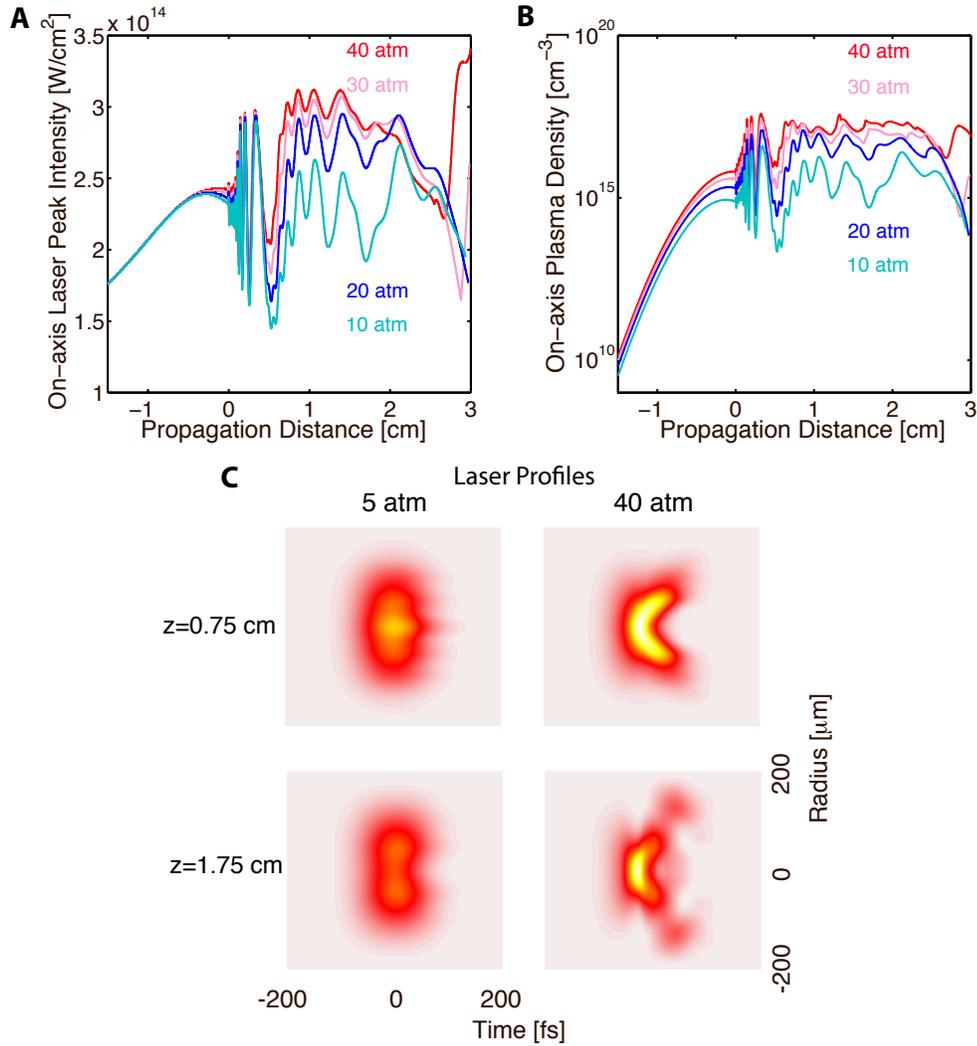

**Fig. S4.** Calculated (**A**) on-axis peak intensity and (**B**) on-axis peak plasma density vs. propagation distance for different He gas pressures. (**C**) Examples of spatio-temporal beam profiles at two different locations for high (40 atm) and low (5 atm) pressures inside the capillary. Here the capillary input is located at $z = 0$. Spatio-temporal confinement is evident after the laser propagates 1.75 cm at high 40 atm pressures inside the waveguide.

Further evidence for laser self-confinement is that phase-matched buildup will occur at a higher pressure for a smaller guided focal spot regardless of whether guiding is extrinsic (i.e. a waveguide) or intrinsic (i.e. self confinement). Thus, in the case of Ar, the second peak in x-ray yield as a function of pressure seen in Figs. 3 and S3 A is expected if the laser beam and x-ray beam size reduces. Figure S4 C also shows that the laser pulse compresses both in time and space as the laser propagates through the long interaction region with high density-length product. This spatio-temporal compression supports phase matching over cm distances, and also greatly reduces any ionization-induced and normal diffraction that might occur at long wavelengths or for shorter interaction lengths.(*38*) Thus, the waveguide geometry is also ideal for laser beam self-confinement and phase matching of the high harmonic generation process in the keV region. Finally, theoretical and experimental results using other gases such as molecular $N_2$ suggest that similar favorable conditions for efficient HHG upconversion in the presence of laser beam self-confinement can be achieved. Thus, remarkably, nonlinear optics and extreme nonlinear optics converge in the x ray region, where both laser self-confinement and HHG phase matching require high pressures and long propagation distances to form a self-guiding laser beam and to extend the quadratic buildup of the HHG signal to compensate for the low single-atom HHG yield.

**Theoretical models of HHG phase matching in the mid-infrared**

Next we present details of our simulations for how the phase-matched HHG emission at keV photon energies builds up over macroscopic distances in a high pressure medium, in order to predict the HHG pulse duration and phase. The kernel of our propagation code is the single-atom harmonic radiator solver based on the strong field approximation (SFA) (*17, 39*) and calculates the HHG radiation field from a single atom taking into account the particular electromagnetic field at the position of the atom. Our code is based on an extension of the strong field approximation (SFA+),(*27*) which increases the quantitative accuracy of the standard SFA. In order to reduce the computing time, the original SFA+ formalism has been simplified using the saddle-point approximation in the plane transverse to the polarization of the laser field for the momentum space integration. This enables us to compute the single-radiator yield within minutes at laser wavelengths in the mid-IR regime, for which an exact solution of the time-dependent Schrödinger equations becomes impractical due to the extremely long computational times required. The ionization and recombination matrix elements within the SFA+ are evaluated for a Helium atom using Roothaan-Hartree-Fock wavefunctions.(*40*) The dipole acceleration is computed from the gradient of the Coulomb potential for He given in (*41*).

Field propagation is then computed from the single atom response using a scheme based on the discrete dipole approximation (*34*) assuming a plane wave incident field, and including ionization, neutrals and group velocity effects in the fundamental field phase. Absorption was taken into account in the propagation of the harmonics. The attosecond pulses have been computed by Fourier transformation of the harmonic spectrum detected on-axis. The target is modeled as Helium gas with a density of $5 \times 10^{19}$ atoms/cm$^3$. The disparity of the temporal scales involved in the computations (femtosecond for the driving field and attosecond for the most energetic harmonics) together with the high density used, requires an extremely precise computation. At present we are able to demonstrate good convergence for propagation distances up to 20 microns, which correspond to the data presented here. In our calculations, the laser pulses are modeled using a sin$^2$ envelope of 1 and 6 cycle FWHM, 3.9 microns wavelength and phase-matching peak intensities of 4.1 and 3.3 x $10^{14}$ W/cm$^2$. The phase matched x-ray spectrum and group delay obtained for the 6 cycle case is shown in Fig. 1C of the main text, as well as the predicted x-ray pulse duration for one of its temporal bursts.

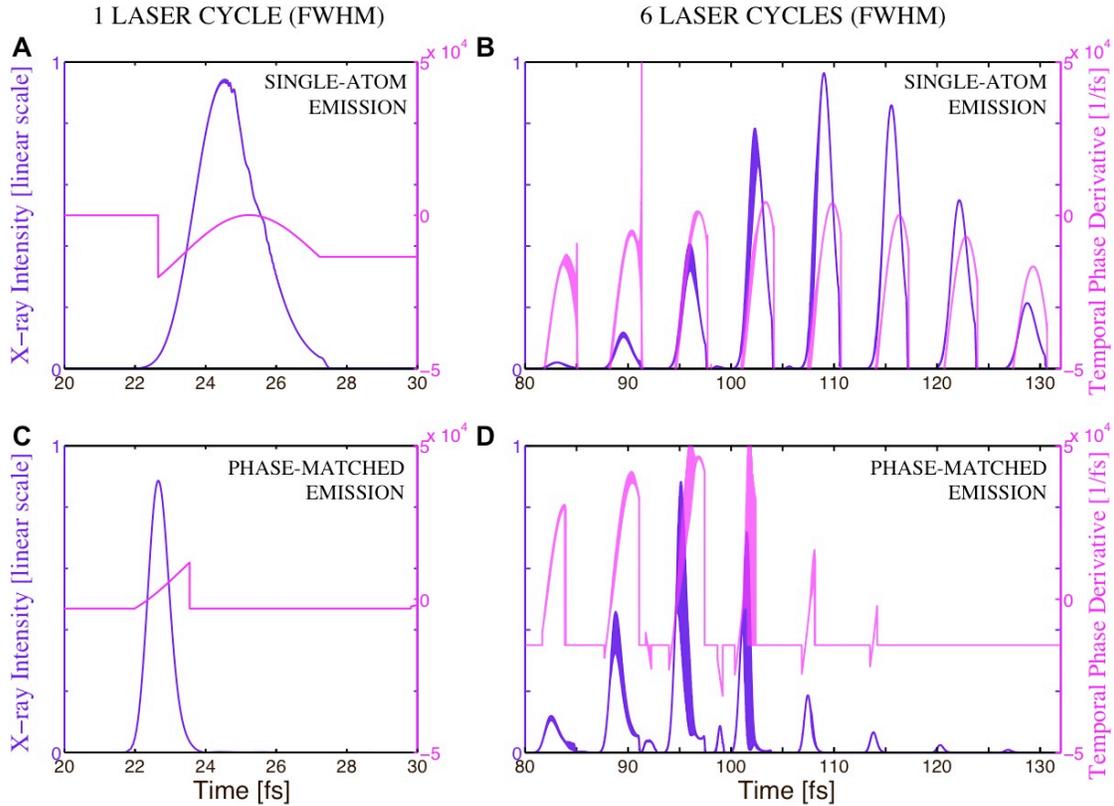

**Fig. S5.** Calculated HHG temporal emission and phase for both a single atom and under phase matching conditions (20 $\mu$m propagation distance) when driven by a 1-cycle FWHM pulse with peak intensity $\simeq 4.1 \times 10^{14}$ W/cm$^2$ (**A** and **C**) and a 6-cycle FWHM (75 fs) 3.9 $\mu$m laser pulse with peak intensity $\simeq 3.3 \times 10^{14}$ W/cm$^2$ (**B** and **D**). The target is He gas with a density of $5 \times 10^{19}$ atoms/cm$^3$. Note that both the duration of the x-ray burst and the number of bursts reduce in the macroscopic phase-matched HHG cases.

A comparison of the HHG temporal emission driven by 1 and 6 cycle FWHM mid-IR

laser pulses, in a phase matched regime, is shown in Figs. S5 C and D. For a 1-cycle FWHM driving laser pulse, Fig. S5 A predicts a burst of 2 fs FWHM HHG burst (associated with ionization due to the first half-cycle of the laser) with a well-behaved chirp. For a 6-cycle FWHM driving laser pulse, the phase-matched HHG emission consists of three intense pulses, each of 2-3 fs duration, also with a well-behaved chirp, and with a total FWHM envelope of 14 fs. The number of x-ray bursts is reduced compared to the single-atom calculation (Fig. S5 B) since the phase-matched emission occurs only over a fraction of the driving laser pulse, near the critical ionization level. As explained in the main text and observed in experiments using 800 nm driving pulses, *(35)* bright HHG emission in the form of an isolated, chirped, x-ray burst is expected for longer propagation distances than those used in Fig. S5, which are at the current limit of advanced theory. The well-behaved chirp means that the x-ray pulse could be compressed in principle, to generate a single-cycle, 2.5 attosecond, Fourier-limited pulse. However, for most time-resolved x-ray experiments, compression to the Fourier limit would not be needed for attosecond time resolution experiments, since no system absorbs over a keV bandwidth. Therefore, knowledge of the chirp would suffice to capture dynamics on multiple atomic sites simultaneously, with attosecond time resolution.

Note that the simulations capture amplitude and phase noise in the x-ray burst, which shifts from the trailing edge to the leading edge of the bursts as the x-rays propagate. Suppressing the higher-order returns of the electron to the ion almost eliminates this noise. However, since emission from higher order returns phase matches under different conditions, this noise is expected to be suppressed for the longer propagation lengths used in the experiment.

In principle, short (positive slope) and long (negative slope) electron trajectories contribute to a temporal chirp. However, in a phase matched waveguide geometry, only short trajectories survive (or in the case of quasi phase matching, either the contribution from the long or short trajectories can be selectively enhanced).(*42*) This is another example of favorable microscopic and macroscopic physics, that in this case selects the short trajectories and reduces phase noise associated with very long trajectories for mid-IR driving wavelengths.

**Analytic model of HHG phase matching from the UV to the keV**

The explicit form of the analytical phase-matching cutoff rule is obtained by inserting the laser intensity at which tunnel ionization occurs,(*43*) which is a function of the critical ionization level, into the single-atom cutoff rule:

$$h\nu_{PM\ cutoff} = I_p + \frac{0.5 I_p^{3+a}}{\ln^2\left[\frac{0.86\tau_L I_p 3^{2n^*-1} G_{lm} C_{n^*l^*}}{-\ln(1-\eta_{CR}(\lambda_L))}\right]} \lambda_L^2$$

where $a=0.5$ is a correction to the analytical approximation, $n^*=Z(13.6/2I_p)^{1/2}$ and $l^*$ are the effective principal and effective orbital quantum numbers, $Z$ is the ionization state, $C_{n^*l^*}=(2e/n^*)^{n^*}/(2\pi n^*)^{1/2}$, where $e$ is the base of the natural logarithms, and $G_{lm}=(2l+1)(l+|m|)!/(6^{|m|}|m|!(l-|m|)!)$. This analytical approximation is valid as long as tunneling ionization does not reach the barrier suppression regime. In barrier suppression ionization, the effective Coulomb potential is bent by the laser field below the valence electron energy.

The exact formula for $\eta_{CR}$ should be used for the phase matching cutoffs, especially for VUV harmonics driven by a short wavelength laser, where the refractive indices of the gas for both the fundamental and HHG wavelengths is rapidly changing. As a result, in the VUV region, the scaling of the phase matching limits deviates from $h\nu_{PM\ cutoff} \propto \lambda_L^{(1.5-1.7)}$, as shown in Fig. 3A.:

$$\eta_{CR} = \left[\frac{\lambda_L^2 r_e N_{atm}}{2\pi\Delta\delta}\left(1 - \frac{1}{q^2}\right) + 1\right]^{-1}$$

However in the soft and hard X-ray regions, our analytic model for phase matching can be simplified, since $\eta_{CR} \ll 1$, and agrees with previous numerical calculations that predict a scaling of the phase matching as $h\nu_{PM\ cutoff} \propto \lambda_L^{(1.5-1.7)}$.(29) In the hard X-ray region of the spectrum, where $\eta_{CR}$ is a tiny fraction of a percent, the analytic phase matching cutoff rule simplifies to:

$$h\nu_{PM\ cutoff} \approx \frac{\alpha I_p^3}{\ln^2[\gamma I_p N \lambda_L^3]} \lambda_L^2$$

where N is the number of laser cycles and taking into account that the critical ionization scales as $\eta_{CR} \propto \lambda_L^{-2}$.